\documentclass[prl, twocolumn,amsmath,amssymb]{revtex4}
\usepackage{notes2bib}

\usepackage{graphicx}
\usepackage{color}

\begin{document}
\title{Colloidal attraction induced by a temperature gradient}

\author{R. Di Leonardo$^1$}
\email{roberto.dileonardo@phys.uniroma1.it}
\author{F. Ianni$^2$}
\author{G. Ruocco$^2$}
\affiliation{$^1$CNR-INFM, CRS SOFT c/o Dipartimento di Fisica, Universit\'a di
Roma ``La Sapienza'', I-00185, Roma, Italy\\
$^2$Dipartimento di Fisica, Universit\'a di Roma ``La Sapienza'',
I-00185, Roma, Italy}

\begin{abstract}

Colloidal crystals are of extreme importance for applied research, such as
photonic crystals technology,  and for fundamental studies in statistical
mechanics.  Long range attractive interactions, such as capillary forces, can
drive the spontaneous assembly of such mesoscopic ordered structures.  However
long range attractive forces are very rare in the colloidal realm.  Here we
report a novel strong and long ranged attraction induced by a thermal gradient
in the presence of a wall.  Switching on and off the thermal gradient we can
rapidly and reversibly form stable hexagonal 2D crystals. We show that the
observed attraction is hydrodynamic in nature and arises from thermal induced
slip flow on particle surfaces. We used optical tweezers to directly measure
the force law and compare it to an analytic prediction based on Stokes flow
driven by Marangoni forces.

\end{abstract}
\maketitle

Colloidal particles, dispersed in a homogeneous solvent, undergo unbiased
Brownian motion unless an external body force, such as gravity, drives them
along a particular direction. Directed colloidal transport can also be induced,
in the absence of a net body force, by a gradient in some external field such
as electric potential (electrophoresis), temperature (thermophoresis) or
concentration of a solute molecule (diffusiophoresis) \cite{anderson}.  In
phoretic transport, due to the absence of a net body force acting on the
particle, and the thin interface layer around it, the surrounding fluid is
perturbed with a flow field decaying as $1/r^3$ , compared with the leading
term of order $1/r$,for sedimentation. As a consequence hydrodynamic
interactions, which are crucial for sedimentation problems, are usually
negligible in phoretic transport. The opposite situation may occur when
particles transport is impeded by an external wall. While sedimenting particles
become stationary as a result of a force balance between gravity and wall
repulsion, phoretic particles have a zero velocity despite the presence of a
non-zero body force (wall repulsion) acting on them. In such peculiar
condition, gradients in the external field will produce a flow field around a
stationary particle which will be long ranged ($1/r$) determining strong
hydrodynamic couplings to nearby particles. Such  hydrodynamic interactions
have been predicted by Squires \cite{squires} and are expected to manifest as
an effective attractive potential when phoretic particles are pinned on a wall.
Here we report the direct observation of strong attraction between colloidal
particles confined at a solid interface by thermophoresis. The induced force is
strong enough to lead to the formation of stable close packed structures and
can be efficiently exploited to drive quick self assembly of colloidal crystal.
Evidences for such a ``hydrodynamic attraction'' in temperature gradients have
been independently reported by Weinert and Braun \cite{braun}. We propose an
analytic expression for the effective force and validate it by direct force
measurement with optical tweezers.

\begin{figure}[]
\includegraphics[width=0.4\textwidth]{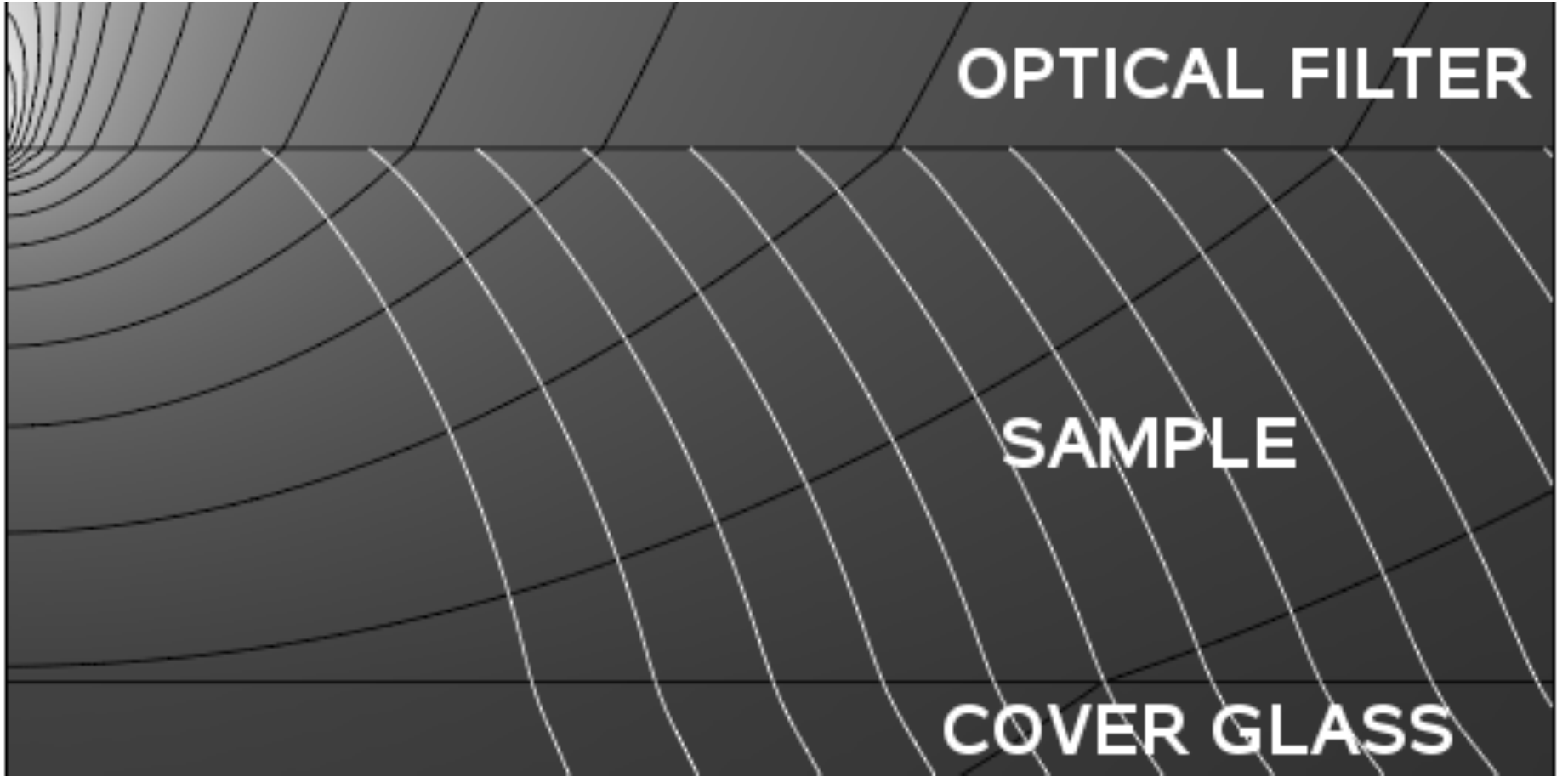}
\caption{
Temperature field in the cell.  Distribution of the isothermal surfaces
(black lines) and the temperature gradient direction (white lines), due to the
absorption of a laser beam by the optical filter at the top.
}
\label{Fig1}

\end{figure}

\begin{figure}[ht]
\includegraphics[width=0.4\textwidth]{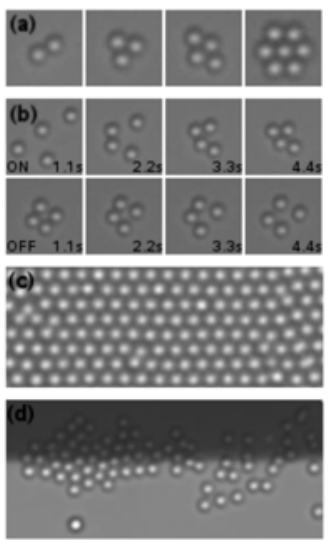}
\caption{
Reversible self-assembly of colloidal particles in a thermal gradient.
a)  Particles confined to the bottom wall by thermophoresis assemble in close-packed structures.
b) Reversible aggregation (first row, temperature gradient on) and disgregation (second row, temperature gradient off)
of closed packed clusters.
c) High concentrated samples assemble in
stable, two dimensional crystal.
d) Close-packed structures form both on a dielectric
charged surface, i. e. the cover glass (visible as the clearer
band), and on a metallic surface (darker band) obtained through a
gold coating 10 nm thick over the glass surface.} \label{Fig2}
\end{figure}

A thin ($\sim$ 15 $\mu$m) sample cell is obtained by sandwiching a small drop
of sample solution (2 $\mu$m diameter silica beads in a glycerol/water mixture)
between a coverslip and an absorbing yellow filter as the microscope slide.
High temperature gradients ($\sim 1 K/\mu m$) can be produced by focusing a
laser beam on the absorbing ceiling of the sample cell (Fig.  \ref{Fig1}). The
same microscope objective (100x NA 1.4 in a Nikon TE2000U inverted microscope)
is used both for focusing the heating beam and for bright field imaging.  As
soon as the laser is turned on, the particles move towards colder regions under
the effect of thermophoresis, following the direction of the negative thermal
gradient, as expected for thermophobic behavior \cite{piazzaparola}.  We used a
finite element method (COMSOL Multiphyics) to solve the stationary heat
equation, with a heat source term modeled as a Gaussian beam, that propagates
in an absorbing medium (top coverslide).  The distribution of the isothermal
surfaces (black lines) and the temperature gradient direction (white lines),
due to the absorption of a laser beam by the optical filter at the top, are
depicted in Fig. 1.  Once the particles reach the bottom wall, a strong
interparticle interaction manifests leading to the formation of closed packed
clusters. In order to reduce the radial components of temperature gradient we
focused on the top absorbing wall an array of 20 traps along a circumference of
30 $\mu$m diameter. Isolated particles in the inner ring region are pushed
against the bottom wall and then perform unbiased 2D Brownian motion. On the
other hand thermophobic particles in the outer region are prevented to enter
the heated region.  In this way we can have a small number of particles in the
inner region and record the dynamics of cluster formation .  Each cluster was
able to find the closest packing structure in a few tens of seconds.  A few
sample clusters are shown in  (Fig. \ref{Fig2}, a).  These clusters are
completely reversible and disgregate as soon as the temperature gradient is
turned off (Fig.  \ref{Fig2}, b).  At higher concentrations particles assemble
in stable, two dimensional crystal (Fig. 3c).

The existence of a confinement-induced long range attraction between particles
of like charge has been observed and debated during the past ten years
\cite{GrierNature,GrierPRE}.  Such interaction is also found to increase when
the like charged particles penetrate the electric double layer at the interface
\cite{GrierNature, goulding}.  One could think that the only role of the
thermal gradient is that of pushing the particles against the double layer of
the bottom interface and that the observed interactions are ultimately
electrostatic in nature.  However like-charge attraction disappears on a
metallic surface \cite{BowenNature,tinoco}.  We then checked (Fig.  \ref{Fig2}
d) that the attraction observed in our experiment persists on a metallic
surface obtained through a gold coating 10 nm thick over the glass surface.
The position of the laser beam inducing the thermal gradient is indicated by
the brighter particle at the bottom, trapped by the beam and displaced from the
bottom wall. Close-packed structures form both on a dielectric charged surface,
i. e. the cover glass (visible as the clearer band), and on a metallic surface
(darker band).  Neither we can attribute the observed attraction to an imaging
artifact \cite{bechinger}, as the formation and stability of small clusters
clearly evidences the presence of a strong attractive force.  Another candidate
for the origin of the attraction could be the distortion of the isothermal
surfaces in the fluid, due to the different thermal conductivity of the
particles and to the spatial asymmetry induced by presence of the wall. As, in
our case, thermophoresis drives particles along the negative thermal gradient,
this effect may induce an attraction between particles touching a colder wall,
when thermal conductivity is higher than the solvent's one, or a repulsion in
the inverse case.  However we always observe attraction in both cases of a
higher (silica) and lower (polystyrene) thermal conductivities.

\begin{figure}[]
\includegraphics[width=0.45\textwidth]{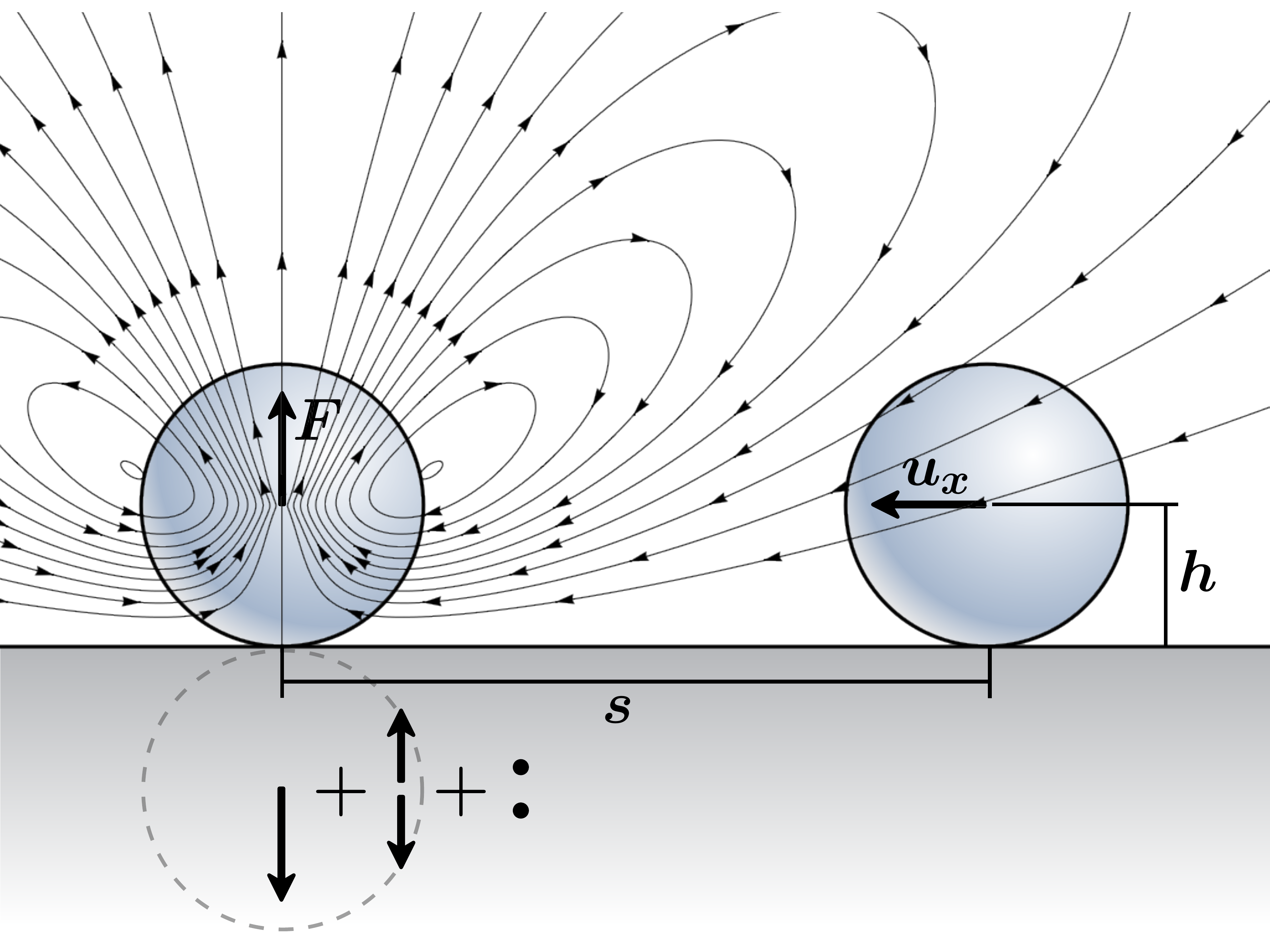}
\caption{Streamlines of the flow generated by a point force pointing away from a plane
wall with sticky boundary conditions. The resulting flow can be represented as the superposition
of the Stokeslet flow produces by the force $F$ plus an image singularity sister consisting
of a Stokeslet, 
a Stokes dipole  and
a source dipole 
\cite{blake}.} \label{stream}
\end{figure}

We suggest, in accordance to \cite{braun}, that the underlying mechanism is
hydrodynamic in nature.  The basic idea is that even if the particles stop when
they reach the bottom wall there's a non zero net force acting on them,
provided by electrostatic repulsion from the wall (the net force and torque
produced by thermal gradients on the interface are always zero
\cite{anderson}). At large enough particle separations the ambient flow
produced by one particle at the location of the other can be obtained as the
flow propagating from a point force. In the absence of the wall such a flow
would be described by the so called Stokeslet \cite{pozri} which wouldn't have
any attractive component on the other particle. When the no-slip condition
imposed by the wall is taken into account, image singularities appear
(Stokeslet, source doublet and Stokes doublet \cite{blake}) contributing a flow
component $u_x$ pulling the the two particles together. The streamlines of the flow
produced by a Stokeslet pointing away from the wall are reported in Fig.
\ref{stream}.  Such a flow will have, on the center of a second particle
touching the wall at a distance $s$, a component parallel to the wall given by
\cite{squires}:

\begin{equation}
u_x=-\frac{3 F}{2\pi\eta}\frac{h^3 s}{\left(s^2+4 h^2\right)^{5/2}}
\label{ux}
\end{equation}

where $F$ is the force exerted by the wall on the particle and $h$ is the
distance between the particle center and the wall.  In the limit of large
interparticle distances $s$ (with respect to the distance $h$ between
particles' centers and the wall), the flow component parallel to the wall
decays as $1/s^4$. The slowly decaying components ($1/s^2$) coming from the
image Stokeslet and Stokes doublet cancel out at any $s$.  Such an ambient flow
on the second particle will be undistinguishable from an attractive force of
intensity:

\begin{equation}
f=6\pi\eta a \lambda u_x=
-9\lambda F \frac{a h^3 s}{\left(s^2+4 h^2\right)^{5/2}}
\label{f}
\end{equation}

where $\lambda=\lambda(a/h)$ is an adimensional correction factor on the Stokes
drag due to the presence of the wall \cite{brenner}.  The particles will then
perform Langevin dynamics under the action of such interaction forces in
addition to the usual drag and stochastic forces. Such an effective force field
is derivable from a potential function and therefore particle statistics will be
governed by Boltzmann statistics with an effective interaction potential:

\begin{equation}
V(s)=-3\lambda F \frac{a h^3}{\left(s^2+4 h^2\right)^{3/2}}
\end{equation}

The force $F$ needed to hold a particle in a temperature gradient can be
evaluated along the same lines suggested by Wurger \cite{wurger} for the
calculation of the thermophoretic speed $\mathbf U$.  In that paper the author
solves Stokes equations for a homogeneous fluid imposing the three boundary
conditions given by: i) a vanishing net force is applied to the particle, ii)
the tangential component of the stresses on the particle surface cancel out the
Marangoni force.  iii) a zero normal component of fluid velocity relative to
the particle surface, which in the lab frame gives:

\begin{equation}
\left. v_r\right|_{r=a}=\mathbf U\cdot\hat{n}
\label{unw}
\end{equation}

In order to satisfy the above conditions one gets a value for the
thermophoretic speed given by \cite{wurger}

\begin{equation}
{\mathbf U}=-\frac{a \gamma_T \kappa}{3 \eta} \boldmath{\nabla} T
\end{equation}

where $\eta$ is the solvent viscosity, $\gamma_T=\partial\gamma/\partial T$ is
the partial derivative with respect to temperature of the 
interfacial energy $\gamma$, and $\kappa=3\kappa_S/(2 \kappa_S+\kappa_P)$ is a
dimensionless combination of particle and solvent thermal conductivities.

We seek here a solution for a spherical particle held stationary in a thermal
gradient by a body force. The corresponding boundary conditions become: i)
again a zero normal component of fluid flow on particle surface in the particle
frame which now coincides with the lab frame, ii) a non vanishing, but unknown
in modulus, net force $F$ on the particle, iii) an off-diagonal stress tensor
canceling the Marangoni force.  A Stokeslet flow has a zero off diagonal stress
tensor, a non vanishing body force $F$, and a fluid flow projection on the
particle surface normal $\hat n$ given by  
\begin{equation}
\left. v_r\right|_{r=a}=\frac{\mathbf F\cdot\hat{n}}{4\pi\eta a}
\label{uns}
\end{equation}

Comparing (\ref{unw}) to (\ref{uns}) it is easy to satisfy the above conditions
by simply adding to the flow obtained above for the freely moving particle, a
Stokeslet flow with 

\begin{equation}
{\mathbf F}=-4\pi\eta {\mathbf U}=\frac{4\pi a^2\gamma_T \kappa}{3}\boldmath\nabla T=\frac{1}{3}\boldmath\nabla \Gamma 
\label{F}
\end{equation}

where $\Gamma$ is the total interfacial energy.  We can now assume that $F$
remains a good estimate even in the presence of the wall and get an expression
for the force $f$ by substituting (\ref{F}) in (\ref{f}). For spheres almost
touching the wall ($h\sim a$ \bibnote{  In taking the limit $h\rightarrow a$ the only singular expression is $\lambda(a/h)$
whose value is therefore the most difficult to estimate.
}) we get:

\begin{eqnarray}
\label{force}
f&=&-f_0 \frac{s/a}{\left[4+(s/a)^2\right]^{5/2}}\\
\label{f0}
f_0&=&12\pi\lambda \kappa \gamma_T\nabla T a^2
\end{eqnarray}

The force depends linearly on the interface energy gradient, increases with the
square of particle size but is independent from solvent viscosity.

In order to validate the above expression we used optical micromanipulation to
directly measure the force law with interparticle distance.

Holographic optical tweezers allow to isolate a single pair of particles and
vary their relative distance while gauging their interaction. In particular, a
high intensity optical trap ($P=42$ mW) is focused next to the interface with
the top wall in order to produce the thermal gradient, while two traps of lower
intensities ($P=3$ mW) are used to hold two particles on the bottom wall. We
checked that the intensity of the last traps alone is low enough not to induce
further attraction between the particles.  In order to avoid fluid
instabilities induced by the high thermal gradient \cite{boon}, we have chosen
a high viscosity suspension (a glycerol-water mixture 56\% w) and a thin cell
gap (13 $\mu m$ thick). 

\begin{figure}[]
\includegraphics[width=0.5\textwidth]{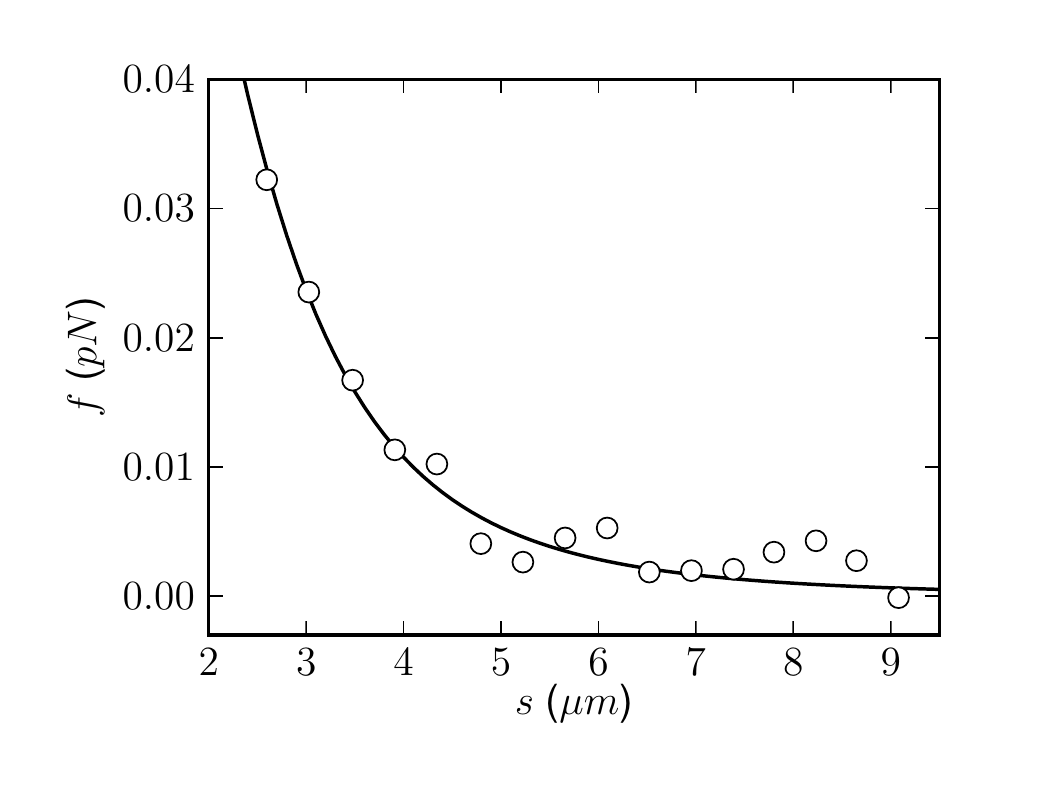}
\caption{
Intensity of the attractive force as a function of
distance. Open symbols are experimental data, solid line is a fit to Eq. (\ref{force}).
}
\label{Fig3}
\end{figure}

The thermal induced interaction tend to push the particles towards each other
and out of the optical traps, until the restoring trap force balances the
attractive force. By measuring the difference $\Delta s$ between trap centers
distance and interparticle distance, the value of the attractive force can be
deduced: $f=k\Delta s$, where $k$ is the equivalent elastic strength of the two
traps system \cite{rdl} and may be determined experimentally from the mean
square displacement of particle distance: $\langle\Delta d^2\rangle=K_BT/k$.
In the reported experimental conditions we found $k=0.1$ pN/$\mu$m.
However an experimental determination of the prefactor $f_0$ in (\ref{f0}) is
very hard because of the strong dependence of $\lambda$ on the exact value of
$h$ and difficulties in measuring the temperature gradient. On the other side
the dependence on interparticle distance only contains particle radius and can
be accurately checked.  Fig.  \ref{Fig3} reports the experimental
determination of force as a function of interparticle distance together with a
fit to Eq. \ref{force} with $f_0=4.8$ pN as the only fit parameter.

In conclusion, we have shown that a thermal gradient pushing thermophoretic
particles against an interface can induce a strong and long ranged attraction
between like-charged particles, leading to the prompt formation of stable,
ordered structures. We provide a static measurement of this interaction and
quantitatively confirm the prediction of a hydrodynamic pseudo-force
\cite{squires} driven by Marangoni forces on particle surface.  The long range
nature of the interaction, similarly to capillary forces, may open new routes
to fabrication of ordered structures.  Already, this mechanism may have
enhanced the self-assembling observed under a convective flow and uniquely
attributed to a confining effect \cite{cheng, toyotama, braunAPL}.

We thank M. Ortolani and the IPN-CNR labs for kindly providing us the gold
coated cover glass.

\bibliography{biblio}

\begin{thebibliography}{20}
\bibitem[{Bibnote1()}]{Bibnote1}
Bibnote1, \bibinfo{note}{in taking the limit $h\rightarrow a$ the only singular
  expression is $\lambda (a/h)$ whose value is therefore the most difficult to
  estimate.}
\end{thebibliography}
\end{document}